\documentclass[prd,twocolumn,groupedaddress,altaffilletter,amssymb,eqsecnum,showpacs,epsfig]{revtex4}
\usepackage[dvips]{graphicx}
\begin{document}
\title{Asymptotic behavior of Cardassian cosmologies with exponential potentials}
\author{Ruth Lazkoz}
\email{wtplasar@lg.ehu.es}
\affiliation{Fisika Teorikoa, Zientzia eta Teknologia Fakultatea, Euskal Herriko Unibertsitatea, 644 Posta Kutxatila, 48080 Bilbao, Spain}
\author{Genly Le\'on}\email{genly@uclv.edu.cu}\affiliation{Universidad Central de Las Villas, Santa Clara CP 54830, Cuba}

% ----------------------------------------------------------------
\vfuzz2pt % Don't report over-full v-boxes if over-edge is small
\hfuzz2pt % Don't report over-full h-boxes if over-edge is small

\date{\today}

\pacs{98.80.-k, 98.80.Jk.}
\begin{abstract}
In this paper we analyze the asymptotic behavior of Cardassian cosmological models filled with a perfect fluid and a scalar field with an exponential potential. Cardassian cosmologies arise from modifications of the Friedmann equation, and among the different proposals within that framework we will choose those of the form $3H^2-\rho\propto \rho^n$ with $n<1$.  We construct two three dimensional dynamical systems arising from the evolution equations, respectively adapted for studying the high and low energy limits. Using standard  dynamical systems techniques we find the fixed points and characterize the solutions they represent. 
We pay especial attention to the properties inherent to the modifications and compare with
the (standard) unmodified scenario. Among other interesting results, we find there are no late-time tracking attractors.
\end{abstract}

%%% ----------------------------------------------------------------------

%%% ----------------------------------------------------------------------
\maketitle
%%% ----------------------------------------------------------------------
\section{Introduction}
Observations indicate that type Ia high redshift supernovae (SNIa) are
dimmer than expected (see \cite{supernovaefirst} for most recent results), and the mainstream
interpretation of this result is that the universe is currently
undergoing accelerated expansion. Further observations, like those of the Cosmic Microwave
Background (CMB) \cite{wmap} or Large Scale structures (LSS) \cite{sdss}, suggest that two
thirds of the energy density of the universe correspond to dark energy.

Even though the most popular explanation to the late-time  acceleration in the universe is the existence of some kind of dark energy (perhaps a scalar field), this is not the only possibility. 
Recently, Freese and Lewis \cite{lewis} proposed the so-called Cardassian models as an alternative explanation  which involves only matter and radiation and does not invoke
either vacuum energy or a cosmological constant. In these models the universe has a flat geometry, as
required by measurements of the cosmic background radiation
\cite{Netter}, and it is filled  only with radiation and matter (baryonic or not). The
Friedman equation is modified with respect to its usual form by the addition of a term in its right hand side, specifically 
\begin{equation}
3H^2=\rho+\sigma\rho^n,
\end{equation} in units such that
${8\pi}/{m_{pl}^2}=1$ and with $\sigma>0$ being an arbitrary constant.

For $n\!<\!1$ the second term becomes important if $z\!<\!\mathcal{O}(1)$. From there on it dominates the Friedmann equation and  yields $a\propto t^{2/3n}$ for ordinary matter,  so there will be acceleration provided $n<2/3$.  There are  two main (possibly unrelated) motivations for the  $\rho^n$ 
modifications: (1) As shown in  \cite{chungfreese}, terms of that form 
typically appear in the Friedman equation  when the universe 
is embedded as a three-dimensional surface (3-brane)
in higher dimensions. (2) Alternatively, these functions may appear
in a purely four dimensional theory in which the modified right
hand side of the Friedman equation is due to an extra contribution
to the total energy density. One will then regard the right hand side of the Friedman equation as corresponding to a single
fluid, with an extra contribution to the energy-density tensor in the
(ordinary four dimensional) Einstein equations.

The interpretation of the Cardassian expansion as due to  
an interacting dark
matter fluid with negative pressure was developed in \cite{gondolofresse}. The Cardassian
term on the right hand side of the Friedman equation is
interpreted as the interacting term and 
gives rise to the effective negative pressure which drives the
cosmological acceleration.

Interestingly, Cardassian models survive several
observational tests,  the most significant being that it allows for a   universe
consisting of just matter and radiation. The
energy density giving a closed universe $\rho_c$ is much smaller that its counterpart
standard cosmology  $\rho_{c \!,\,\rm{old}}$ (specifically
$\rho_c=\rho_{c\!,\,\rm{old}}\left[1+(1+z_eq)^3(1-n)\right]^{-1}$),
 and matter alone is enough to provide a
flat geometry.

With regard to the observational tests that depend only on the scale or the Hubble factor, in the late-time regime Cardassian models filled with just matter ($ \rho\propto a^{-3}$)
are indistinguishable from perfect fluid models with a $p=(\gamma-1)\rho$ equation of state
under the identification $n\equiv \gamma$. These perfect fluid models are in turn 
kinematically equivalent to scalar field (quintessence) models with an exponential potential. In this way, the Cardassian model can make
contact with quintessence with regard to observational tests. Interestingly, observational 
tests seem to favor $n<0$, so that asymptotically one would get a phantom equation of state \cite{phantom}.

However, the equivalence between Cardassian and perfect fluid models
is not extensible to the dynamical realm, the evolution of perturbations may differ significantly, and this can lead to discrepancies for instance in observational tests
associated with the cosmic microwave bakground. 

Nevertheless, as stated in \cite{astro-ph/0201229}, questions of interpretation
remain open, because  in the Cardassian model matter alone is responsible for the accelerated
behavior, and yet the universe can be flat.

Here we perform a dynamical systems analysis of Cardassian models. These techniques have 
been used for exploring  Cardassian models filled
with baryonic matter only \cite{cardin}, but we give one step further  by allowing as well for a scalar field component
(non-baryonic matter). We investigate  for the early and late time regime of the models the existence of tracking solutions, i.e., we look for solutions representing  universes with  non-negligible and proportional fractions of both baryonic and non-baryonic matter. 
Tracking solutions are particularly interesting because their dynamical effects mimic a decaying cosmological constant (see \cite{copeland,seminal} for seminal references). Such solutions would be devoid of  the fine-tuning problems posed by a cosmological constant precisely because of the independence on the initial conditions. 

Using standard dynamical systems techniques we will investigate the interplay between the modifications to the Friedmann equations and the features and existence conditions of the mentioned tracking
solutions.
In this first approach to the problem we will choose an exponential self-interaction potential \cite{exp,copeland,seminal} for the scalar field because only those potentials allow using the evolution equations for constructing an autonomous system of as low a dimension as possible. As shown in \cite{coptrack}, for any other potential the number of dimensions will be higher if the system is to remain autonomous. In that reference the existence of tracking Cardassian cosmologies was shown within a more general context, but their analysis is not as detailed as ours. 

The remark has to be done that even though observations seem to favor $n<0$ \cite{favor}, we will restrict our numerical examples to cases with $0<n<2/3$ so that we can establish comparisons between our results and those in \cite{copeland}. Nevertheless, most aspects of our analysis do not depend on the sign of $n$, but just on the requirement that $n<1$.

The plot of the paper is as follows. In Section II we study the phase-space, find its fixed points and characterize them. In Section III we discuss the cosmological consequences of the attractor solutions. Finally, in Section IV we outline our main conclusions, including
our main result that there are no tracking late-time attractors.

 \section{Phase-space}
The evolution equations for a flat Friedmann-Robertson-Walker (FRW) Cardassian cosmological
model filled with a scalar field $\phi$ with self-interaction potential $V(\phi)=\xi\exp(-\sqrt{6}\lambda\phi)$ and a barotropic perfect fluid with equation of state $p_{\gamma}=(\gamma-1)\rho_{\gamma}$ are
\begin{eqnarray}
&&2\dot H+\left(\gamma\rho_{\gamma}+\dot\phi^2\right)
\left(1+n\sigma \rho_{\rm tot}^{n-1}\right)=0\label{F2}
\\
&&\ddot\phi+3H\dot\phi+\frac{d V(\phi)}{d\phi}=0,\label{KG}\\
&&\dot\rho_{\gamma}+3\gamma H\rho_{\gamma}=0 ,\label{CE}
\end{eqnarray}
where for the total energy density $\rho_{\rm tot}$ we have
\begin{equation}
	\rho_{\rm tot}=\frac{1}{2}\dot\phi^2+V(\phi)+\rho_{\gamma}.
\end{equation}

The evolution equations (\ref{F2}-\ref{CE}) are in turn subject to the constraint
\begin{equation}
	H^2=\frac{1}{3}\rho_{\rm tot}\left(1+\sigma \rho_{\rm tot}^{n-1}\right).\label{F1}\\
\end{equation}
Here and throughout
$\sigma$, $k$  and $\lambda$ will be free parameters, and we will restrict ourselves
to the $n<1$ case.

Experience has demonstrated that dynamical systems methods can be used to
describe the evolution of cosmological models  by means of 
past and future attractors.
In order to cast our set of equations as a dynamical system, it is convenient 
to normalize the variables,  because in the vicinity of an hypothetical initial singularity
physical variables would typically diverge, whereas at late times they commonly
tend to zero \cite{wainlim}. Due to physical considerations
normalization with the Hubble
factor is an appropriate choice in cosmology.
Besides, all available mathematical evidence suggests that Hubble-normalized
variables are bounded into the past (that is, as the initial
singularity is approached), and if there is a cosmological constant (or something that mimics it) it seems those variables will also be bounded into the future.
Thus, even though the Hubble-normalized state space is unbounded, it is sensible to expect
that the evolution equations will admit a past attractor and a future attractor.
\subsection{High energy limit}
In view of the discussion above, and following the usual practice, we use normalized (dimensionless) variables defined by
\begin{eqnarray}
&&x=\frac{\dot\phi}{\sqrt{6}H}\label{v1x},\\
&&y=\frac{\sqrt{V}}{\sqrt{3}H}\label{v1y},\\
&&z=\frac{\sigma^{\frac{1}{1-n}}}{3H^2}\label{v1z},\\
&&w=\frac{\sqrt{\rho_\gamma}}{\sqrt{3}H}\label{v1w}.
\end{eqnarray}

This coordinates will allow us analyzing the solutions of (\ref{F2}-\ref{KG}), and the cosmological models associated with them.
As will become clear below, this choice of coordinates  leads to a dynamical system which is best adapted for studying high energy limit solutions, so we will refer to it as our high energy limit adapted (HELA) dynamical system as
opposed to a low energy limit adapted (LELA) that will emerge below after an alternative choice of coordinates.

In addition, the variables will be related among them through
\begin{equation}
\mathcal{S}(w,x,y)+\mathcal{S}^n (w,x,y)z^{1-n}=1\label{lig},
\end{equation}
where $\mathcal{S}(w,x,y)=w^2+x^2+y^2.$

The constraint (\ref{lig}) lets us ``forget'' about the evolution of one of the coordinates. Here we will choose the discarded coordinate to be $z$. Using the variables (\ref{v1x}-\ref{v1w}), equation (\ref{lig}), and the conservation equations
(\ref{KG}) and (\ref{CE}) we get the
equations
 {\setlength{\arraycolsep}{0pt}
\begin{eqnarray}
x'&=&\frac{1}{2\mathcal{S}}\left(x\left(
2\left( n -1\right) x^2 - 2y^2 + \left(\gamma  n  -2\right) w^2 \right)- 
   \right.
\nonumber\\&&\left.\left( \left( n-1\right)x\left(2 x^2 +\gamma xw^2  \right)-
     2\lambda y^2  \right)\mathcal{S}
\right), \qquad\label{eq1x}\\\nonumber\\
y'&=&\frac{y}{2\mathcal{S}}
\left(n\left(2x^2 +\gamma w^2\right)  - 
\right.\nonumber\\&&\left.
\left(2\lambda x+\left( n-1 \right)(2x^2 + 
     \gamma w^2)\right)\mathcal{S}
\right),\nonumber\\
\label{eq1y}\\
w'&=&\frac{w}{2\mathcal{S}}
\left(
x^2\left( 2n - \gamma  \right)  + \gamma\left( n -1\right) w^2  -  \right.\nonumber\\
&&\left.\gamma y^2  -
\left( n -1\right) (2 x^2 + \gamma w^2)\mathcal{S}
\right),\label{eq1w}
\end{eqnarray}}

\noindent where primes denote differentiation with respect to  $\tau\equiv\ln a^3.$   Equations ({\ref{eq1x}-\ref{eq1w})  form the HELA dynamical system, which is
defined on the  phase space
\begin{equation}
{\Psi}=\left\{(x,y,w):\mathcal{S}\le1\right\}.
\end{equation}

To begin with the study of the evolution of our the dynamical system  we have to find its fixed (or critical) points. The fixed points $(x^{\star},y^{\star},w^{\star})$ are given by the conditions
\begin{eqnarray}
 x'\left(x^{\star},y^{\star},w^{\star}\right)=0, \\
  y'\left(x^{\star},y^{\star},w^{\star}\right)=0 ,\\
  w'\left(x^{\star},y^{\star},w^{\star}\right)=0. 
\end{eqnarray}
In order to analyze the stability of the fixed points $(x^{\star},y^{\star},w^{\star})$  one 
studies the linearized dynamical system obtained by expanding
equations (\ref{eq1x})--(\ref{eq1y}) about those fixed points (see, e.g \cite{hirsch}). 
Then one tries solutions in the form $(x,y,w)=(c_1,c_2,c_3)\,e^{\lambda t}$ in the linear
approximation, and finds that the characteristic exponent $\lambda$ and the constant
vector $(c_1,c_2,c_3)$ must be respectively an eigenvalue and an eigenvector of the matrix
\begin{equation}
\left( \begin{array}{ccc}
\displaystyle\frac{\partial x'}{\partial x}&   
\displaystyle\frac{\partial x'}{\partial y}
&   
\displaystyle\frac{\partial x'}{\partial w}\\
   \displaystyle\frac{\partial y'}{\partial x} &   \displaystyle\frac{\partial y'}{\partial y} &   \displaystyle\frac{\partial y'}{\partial w} \\
   \displaystyle\frac{\partial w'}{\partial x} &   \displaystyle\frac{\partial w'}{\partial y} &   \displaystyle\frac{\partial w'}{\partial w}
\end{array} \right)_{(x,y,w)=(x^{\star},y^{\star},w^{\star})}.
\end{equation}

\begin{widetext}
\begin{center}
\begin{table}[t!]\begin{center}\caption{\label{locHELA}Location and existence conditions of the critical
points of the HELA dynamical system}
\end{center}
\begin{tabular}{cccccccc}
\hline \hline
Name &$x$&$y$&$w$&\quad Existence\quad& \\
\hline\vspace{5pt}
$\,W_{\pm}\,$ &$0$&$0$&$\pm1$&\quad All $\gamma$ and $\lambda$\quad& \\\vspace{5pt}
$\,X_{\pm}\,$ &$\pm1$&$0$&$0$&\quad All $\gamma$ and $\lambda$\quad& \\\vspace{5pt}
$XY_{\pm}W_{+}\,$
&$\quad\displaystyle\frac{\gamma}{2\lambda}\quad$&$\quad\pm\displaystyle\frac{1}{2\lambda}\sqrt{{\displaystyle(2-\gamma)\gamma}}\quad$
&$\quad\sqrt{1-\displaystyle\frac{\gamma}{2\lambda^2}}\quad$&\quad   $\quad\displaystyle 1\ge\lambda^2\geq\frac{\gamma}{2}\quad$\quad& \\\vspace{5pt}
$XY_{\pm}W_{-}\,$
&$\;\displaystyle\frac{\gamma}{2\lambda}\;$&$\,\pm\displaystyle\frac{1}{2\lambda}\sqrt{{\displaystyle(2-\gamma)\gamma}}\,$
&$\;-\sqrt{1-\displaystyle\frac{\gamma}{2\lambda^2}}\;$& $\quad\displaystyle 1\ge\lambda^2\geq\frac{\gamma}{2}\,$\quad& \\\vspace{5pt}
$XY_{\pm}\,$&$\lambda$&$\pm\sqrt{1-\lambda^2}$&$0$&\quad $\lambda^2\leq 1$\quad&
\\
$\rm{O}\,$&$0$&$0$&$0$&\quad All
$\gamma$ and $\lambda$\quad&
\\
\vspace{-0.30cm}\\
\hline \hline
\end{tabular}
\end{table}
\end{center}
\end{widetext}

If the real part of the three characteristic exponents is negative, the
fixed point is asymptotically stable, i.e., an attractor. On the other
hand, it is enough to have (at least) one characteristic exponent with
positive real part to make the fixed point asymptotically unstable, i.e., a repeller. This
repeller is a saddle point if at least one of the other characteristic
exponents has a negative real part, in which case there is, apart from the
unstable manifold, a stable manifold containing the exceptional orbits
that converge to the fixed point. The case in which the largest real
part is precisely zero must be analyzed  using other methods, for the
linear analysis in unconclusive. The geometric form of the orbits near
the fixed point is determined by the imaginary part of the
characteristic exponents. If the three are real the fixed point is a
node. A couple of complex conjugate exponents leads, except in
degenerate cases, to an spiral point (the orbits are helices near the
fixed point). In addition, when one of the exponents is null the point is not hyperbolic
and therefore structural stability cannot be guaranteed (the geometric form of the trajectories may change under small perturbations).
\begin{widetext}
\begin{center}
\begin{table}[h!]
\caption{\label{eigenHELA} Eigenvalues and dynamical character  of the critical points of the
HELA dynamical system}
$\big( $assuming $n<1$, $0\le\gamma\le2$ and $\lambda^2\le1$ and $\beta=\sqrt{(2-\gamma)(4\gamma^2+\lambda^2(2-9\gamma))}$ $\big)$\vspace{0.3cm}
\begin{tabular}{ccc}
\hline \hline
Name &\quad Eigenvalues &\quad Dynamical Character\\\\
\hline\\
$W_{\pm}$ &\quad
$\left(\displaystyle\frac{1}{2}(\gamma-2),\;\frac{\gamma}{2},\;\gamma(1-n)\right)$&
non-hyperbolic for $\gamma=0,2$, repeller for $0<\gamma\le2$  (saddle for $0<\gamma<2$)\\
\\
$X_{\pm}$ &\quad
$\left(2(1-n),\;1-\displaystyle\frac{\gamma}{2},\;1\mp\lambda\right)$ &non-hyperbolic for  $\gamma=2$ or $\lambda=1$, always repeller   \\\\
$XY_{\pm}W_{+}$&\quad $\left(\gamma(1-n),\;
\displaystyle\frac{(\gamma-2)\lambda-\beta}{4\lambda^2},
\;\displaystyle\frac{(\gamma-2)\lambda+\beta}{4\lambda}\right)\quad$&\quad
non-hyperbolic for $\gamma=0,2$, repeller for $\gamma>0$ \\
&&(saddle or spiral-out saddle for $0<\gamma<2$)\\
\\
$XY_{\pm}W_{-}$ &\quad $\left(\gamma(1-n),\;
\displaystyle\frac{(\gamma-2)\lambda-\beta}{4\lambda},
\;\displaystyle\frac{(\gamma-2)\lambda+\beta}{4\lambda}\right)$&\quad
non-hyperbolic for $\gamma=0,2$, repeller for $\gamma>0$\\
&&(saddle or spiral-out saddle for $0<\gamma<2$)\\
\\$XY_{\pm}$&\quad
$\left(\lambda^2-1,\;\lambda^2-\displaystyle\frac{\gamma}{2},\;2(1-n)\lambda^2\right)$&\quad
non-hyperbolic for $\lambda^2=1,\gamma/2$, always repeller\\
&&(saddle  for $\lambda^2<1$ or $\lambda^2<\gamma/2$)\\
\\
$\rm O$ &$(0,0,0)$& undefined\\\vspace{-0.30cm}\\
\hline \hline
\end{tabular}
\end{table}
\end{center}
\end{widetext}

Depending on the values of $\lambda$ and $\gamma$ the HELA system above has up to eleven critical points. We will denote them as 
$W_{\pm},\;X_{\pm},\;XY_{\pm}W_{-},\;XY_{\pm}W_{+},\;XY_{\pm},\rm{O}$, and the first ten are closely related those found in Ref. \cite{copeland}, where the $n=0$ case of our system was studied.

 Observe that the equations are invariant under the variable changes
$y\to -y$, $z\to -z$, but not under $x\to
-x$, and so in our numerical examples we will concentrate on the region  $\{x^2+y^2+z^2\geq1,
-1\leq x\leq 1,y\geq 0,w\geq 0\}$, which is a fourth of the unit sphere. This is equivalent to saying we are just considering expanding universes ($H>0$). However, an analytic description of all eleven points is
presented in the lines below and in tables (\ref{locHELA}) and (\ref{eigenHELA}). We will also set restrictions $0\le\gamma\le2$ and $\lambda\le1$ so that neither the barotropic fluid nor 
the scalar field have supraluminical sound speeds and the fluid satisfies the weak energy condition. We will also assume $n<1$, as this is the case of interest.

Although the location of the critical
points of this dynamical system does not depend  on $n$ \footnote{This is confirmed by numerical tests.}, the same is not true for their dynamical character. Note as well that all the critical points but the point $\rm{O}$ are located on the hypersurface $\mathcal{S}=1$, that is, they correspond to $z=0$ and $H=\infty$,  which in turn implies those are solutions to be obtained under the high energy limit approximation. Interestingly, the fact that $H=\infty$ means those are singular (big-bang) cosmologies. 

 In what follows, and in order to complete the information provided in the tables, we will characterize the cosmological models represented by the fixed point living in the above mentioned fourth of the unit sphere.

The first point, called $W_+$,  represents a solution completely dominated by the fluid. The unstable character of these solutions agrees with what  one might have anticipated, are they are only
expected to be relevant at early times.

The second point, called $X_+$,  represents a solution completely dominated by the scalar field, more specifically by its kinetic energy. Figs.  \ref{first} and \ref{second} strongly suggest it is a past attractor.

The third point, called $XY_+W_+$, depicts a tracking  solution, neither the fluid nor the scalar field dominate completely \cite{wet}.  
 
The fifth point, called $XY_+$, represents a  scalar field dominated solution, which is inflationary if $\lambda^2<1/3$
\cite{olkit}.  Figs.  \ref{third} and \ref{fourth} show it is a unstable saddle in the asymptotic future.

As for the sixth point, called $\rm O$ the linear analysis is not sufficient for determining its dynamical character, because it has three null eigenvalues. One possibility is resorting to numerical inspection. As shown in the figures below, this approach tell us the fifth point is the late time attractor of the system. Another tricky aspect of the analysis of this point is the sort of solutions it represents. Clearly, in the vicinity of $(0,0,0)$ we have solutions in which $\dot\phi\ll H$, $V\ll H$ and $\rho_\gamma\ll H^2$, but nothing can be said about the ratios $\dot\phi^2/V$, $V/\rho_\gamma$ and $\dot\phi^2/\rho_\gamma$. This means we cannot say whether the solution  represented by $\rm O$  is scalar field dominated,  fluid dominated  or tracking. Note as well that in the vicinity of  $\rm O$ we have ${\cal S}^n(w,x,y)
z^{1-n}\simeq 1$, which in turn implies $3H^2\simeq \sigma\rho_{\rm tot}^n$, that is, the points in that region represent solutions in which the corrections terms in the Friedmann equation dominate, and this obviously happens in the low energy regime. We have seen this system is good
for the description of the points far away from that region (i.e. associated with solutions in the high energy regime), so  we  say the system is adapted to a high energy limit description and thus dub it HELA system, as opposed to the system adapted to a low energy limit description which will be used in the next subsection, which we dub LELA system. By
reformulating the whole analysis using the LELA system as an alternative to the HELA one, we will be able to unveil the fine structure of the $\rm O$ point, and more importantly we will be allowed to tell what sort of solution is the late-time attractor.

\begin{figure}[t!]
\begin{center}
\hspace{0.4cm}\includegraphics[width=7cm,height=4.5cm]{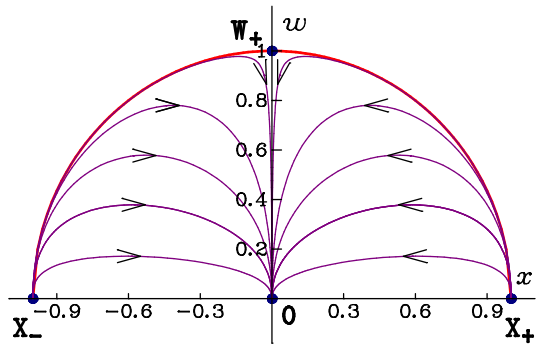}
\caption{Phase-space trajectories lying on the $y=0$ hypersurface
for $\gamma=1$, $n=0.2$ and $\lambda=1$. }
\label{first}
\end{center}
\end{figure}
\begin{figure}
\begin{center}
\hspace{0.4cm}\includegraphics[width=7cm,height=4.5cm]{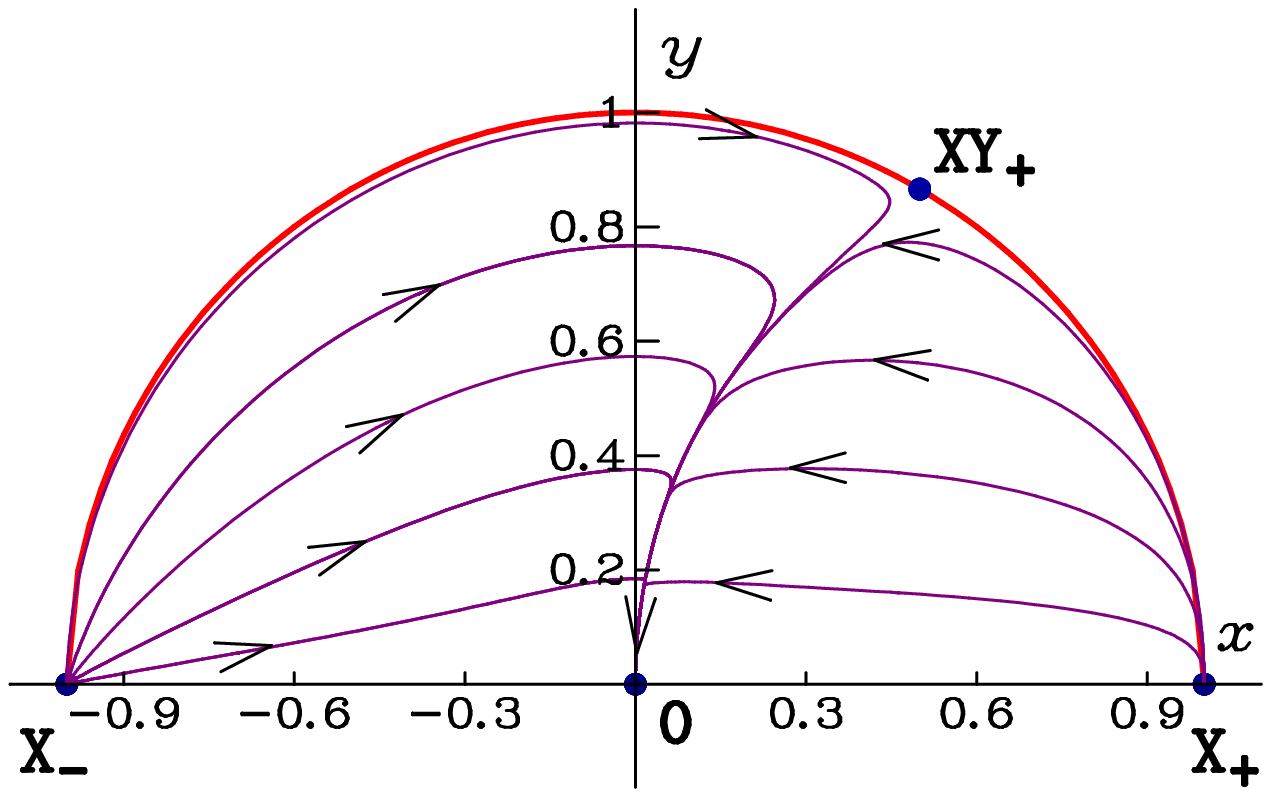}
\caption{Phase-space trajectories lying on the $w=0$ hypersurface
for $\gamma=1$, $n=0.2$ and an arbitrary $\lambda$. }
\label{second}
\end{center}
\end{figure}

\subsection{Low energy limit}
We consider now dimensionless variables defined by
\begin{eqnarray}
&&x=\frac{\sqrt{2}}{2}\sigma^{\frac{1}{2n}}\left(\sqrt{3}H\right)^{-\frac{1}{n}}\dot
T,\label{v2x}\\
&&y=\sigma^{\frac{1}{2n}}\left(\sqrt{3}H\right)^{-\frac{1}{n}}\sqrt{V},\label{v2y}\\
&&w=\sigma^{\frac{1}{2n}}\left(\sqrt{3}H\right)^{-\frac{1}{n}}\sqrt{\rho},\label{v2w}\\
&&z=\sigma^{\frac{1}{2n(1-n)}}\left(\sqrt{3}H\right)^{-\frac{1}{n}},\label{v2z}
\end{eqnarray}
where the above variables are related by

\begin{equation}
\mathcal{S}(w,x,y)^n+\mathcal{S}(w,x,y)z^{-2(1-n)}=1\label{lig2}.
\end{equation}

\begin{figure}[t!]
\begin{center}
\hspace{0.4cm}\includegraphics[width=7cm,height=4.2cm]{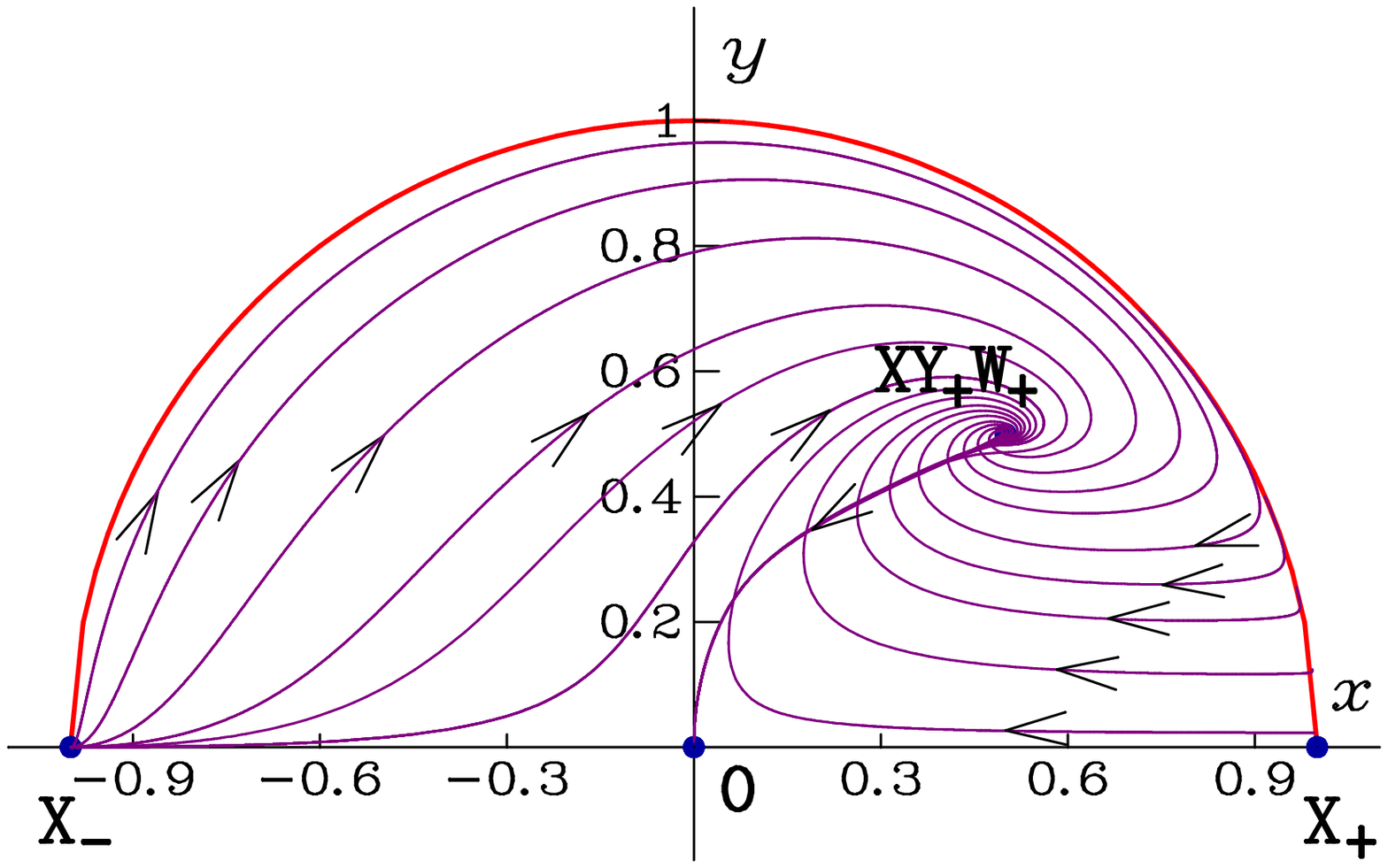}
\end{center}
\caption{XY projection of some phase-space  trajectories for
$\gamma=\lambda=1$ and $n=0.2$. All trajectories cross the
hypersurface $w=0.69$. }
\label{third}
\end{figure}

\begin{figure}[t!]
\begin{center}
\hspace{0.4cm}\includegraphics[width=7cm,height=4.5cm]{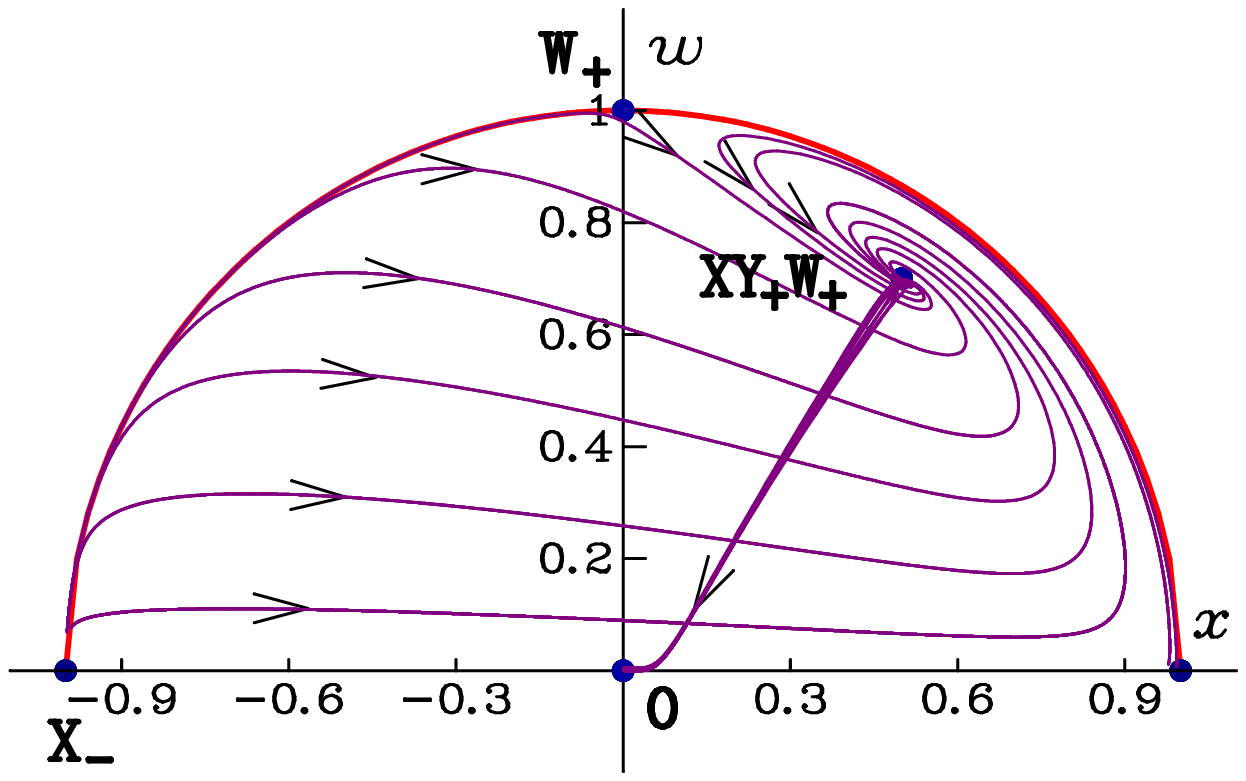}
\end{center}
\caption{XW projection of some phase-space  trajectories for
$\gamma=\lambda=1$ and $n=0.2$. All trajectories cross the
hypersurface $w=0.69$. }
\label{fourth}
\end{figure}

 Using variables(\ref{v2x}-\ref{v2z}) ,
 the relation (\ref{lig2}), and the field equations
(\ref{CE}-\ref{F1}) we get the
equations
{\setlength{\arraycolsep}{0pt}
\begin{eqnarray}
x'=&&\frac{1}{2nS}\left(
2\left(  n -1\right)\left( S^n-1 \right) x^3 - 2nxy^2 + \right.\nonumber\\&&\left.
  w^2x\left(\left( 1 + \left( n -1\right) S^n \right) \gamma -2n \right)  + 
  \right.\nonumber\\&&\left.
  2n{\sqrt{S}}{\sqrt{1 - S^n}}y^2\,\lambda
\right),\nonumber\\\label{eq2x}\\
y'=&&\frac{y}{2nS}\left(
\left( 1 + \left( n-1 \right)S^n\right) \left( 2x^2 + \right.\right.
\nonumber\\
&&\left.\left.\gamma w^2 \right)  - 
  2n\lambda x{\sqrt{S}}\,{\sqrt{1 - S^n}}\right)\label{eq2y},\\
w'=&&\frac{1}{2nS}\left(\left(n-1 \right) \left( S^n -1\right)w^3\gamma  - 
\right.\nonumber\\&&\left.
  w\left( n\gamma (y^2-1)  -
     x^2\left( 2\left( 1 + \left( n -1\right)S^n \right)   \right) 
     \right)\right)\label{eq2w}.
\end{eqnarray}}

Equations ({\ref{eq2x}-\ref{eq2w})  form a dynamical system
defined on the  state space 
\begin{equation}
{\Psi}=\left\{(x,y,w):\mathcal{S}(w,x,y)\le1\right\}.
\end{equation}
As discussed above this will be our LELA system, and like in the HELA one,
the location of the critical
points of this dynamical system  does depend  on $n$, but their the dynamical character does.

\begin{center}
\begin{table}[b!]\caption{Location and existence conditions of the critical points 
of the LELA system}
\begin{tabular}{ccccc}
\hline \hline
Name & $x$ & $y$ & $z$ & Existence\\
\hline \vspace{5pt}
$X_{\pm}$ & $\pm1$ & $0$ & $0$ & All $\gamma$ and $\lambda$
\\\\
$Y_{\pm}$ & $0$& $\pm 1$ & $0$ &All $\gamma$ and $\lambda$
\\\\
$W_{\pm}$ & $\quad 0\quad$& $\quad\pm 1\quad$ & $\quad 0\quad$ &All $\gamma$ and $\lambda$ \\\\
$\rm O$ & $\quad 0\quad$& $\quad 0\quad$ & $\quad 0\quad$ &All $\gamma$ and $\lambda$ \\
\vspace{-0.30cm}\\
\hline \hline
\end{tabular}
\label{critB}
\end{table}
\end{center}

The LELA system has always seven critical points
$W_{\pm},\; X_{\pm},\; Y_{\pm},\;\rm{O}$, no matter the values of $\lambda$ and $\gamma$. Once again, given the invariance under the transformations
$y\to -y$, $z\to -z$, combined with the lack of invariance under the transformation $x\to
-x$, we will concentrate on the region  $\{x^2+y^2+z^2\geq1,
-1\leq x\leq 1,y\geq 0,w\geq 0\}$ in our numerical examples, and as explained above this will mean we will address   expanding  universes only. However, an analytic description of all seven points is
presented in the lines below and in tables \ref{critB} and \ref{eigenB}. We will also set the same restrictions on $\gamma$, $\lambda$ and $n$ as above.

\begin{figure}[t!]
\begin{center}
\hspace{0.4cm}\includegraphics[width=7cm,height=4.5cm]{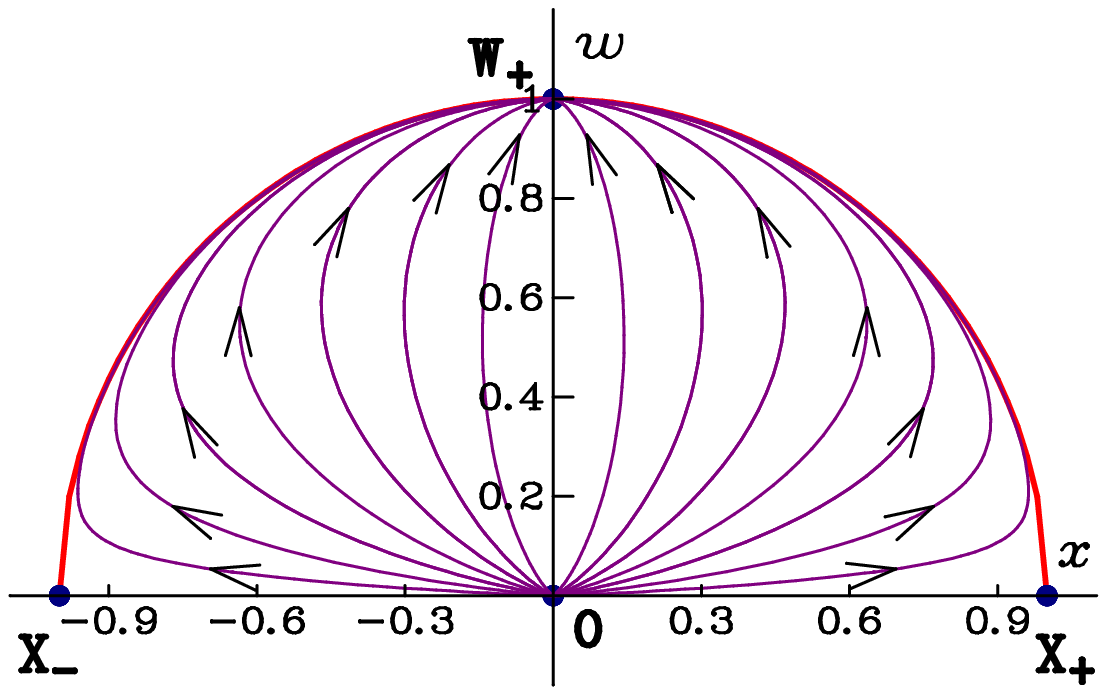}
\caption{Phase-space trajectories lying on the $y=0$ hypersurface
for $\gamma=1$, $n=0.2$ and $\lambda=1$. }
\hspace{0.4cm}\includegraphics[width=7cm,height=4.5cm]{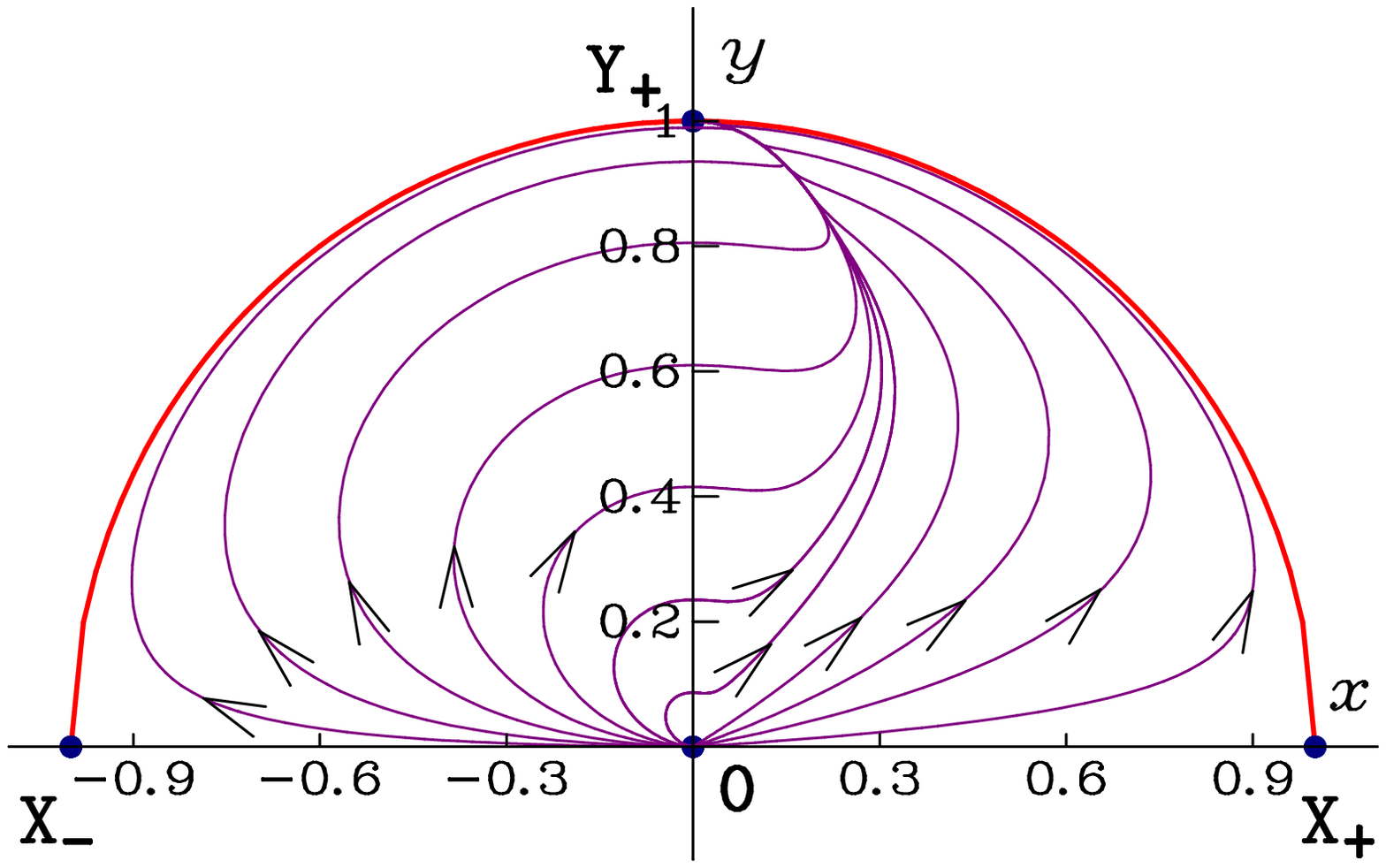}
\end{center}
\caption{Phase-space trajectories lying on the $w=0$ hypersurface
for $\gamma=1$, $n=0.2$ and an arbitrary $\lambda$. }
\label{sixth}
\end{figure}

\begin{figure}
\hspace{0.4cm}\includegraphics[width=7cm,height=4.2cm]{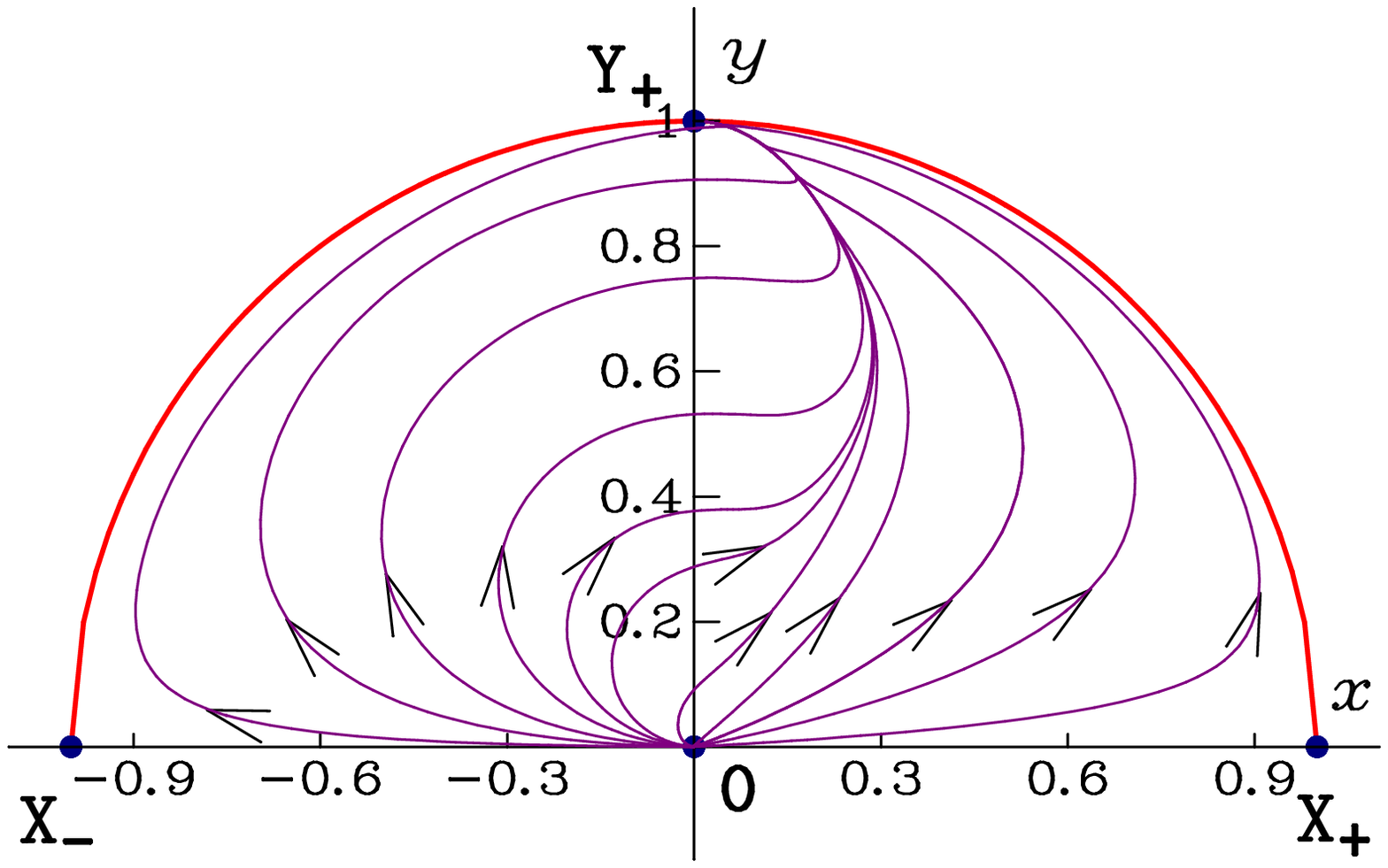}
\caption{XY projection of some phase-space  trajectories for
$\gamma=\lambda=1$ and $n=0.2$. All trajectories cross the
hypersurface $w=0.1$. }
\label{seventh}
\end{figure}

\begin{figure}
\hspace{0.4cm}\includegraphics[width=7cm,height=4.5cm]{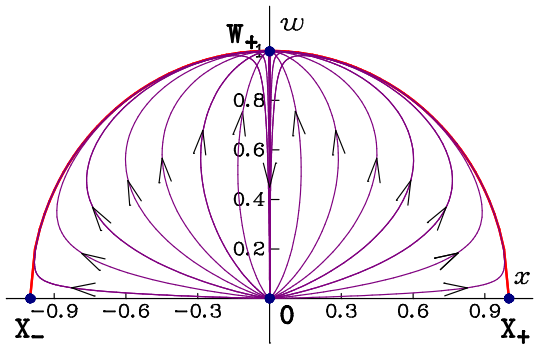}
\caption{XW projection of some phase-space  trajectories for
$\gamma=\lambda=1$ and $n=0.2$. All trajectories cross the
hypersurface $y=0.01$. }
\label{eighth}
\end{figure}
We have seen that the HELA system is not suitable for the description of solutions satisfying $3H^2\approx \sigma\rho^n$. Under the definitions used to construct our LELA system such solutions live in the  region  $\mathcal{S}\approx 1$,  so the solutions represented by the fixed points of $W_{\pm},\; X_{\pm}$ and $Y_{\pm}$ of our LELA system
are also solutions where corrections are important. From  (\ref{lig2}) we see those points are characterized by $z=\infty$, which is a condition  equivalent to $H=0$. Since, in an expanding
universe $H$ goes to zero at late times, the previous conclusion is consistent with the
requirement that at late times acceleration appears due to the domination of  corrections.

In contrast the fixed point $\rm{O}$ represents solutions for which $3H^2\approx \rho$, that is, solutions in a regime in which corrections are negligible, but since these solutions were already investigated during our analysis of the HELA system \footnote{There they were called $W_{\pm},\;X_{\pm},\;XY_{\pm}W_{-},\;XY_{\pm}W_{+},\;XY_{\pm}$.}, we will not
bother to study the features of this $\rm{O}$ point.

In such limit, the critical points  are  $X_{\pm},$ $Y_{\pm},$ and
$W_{\pm},$
 In (\ref{critB}) we summarize the information
about these points.

The first point, called $X_+$, is located at $(x,y,w)=(1,0,0)$. It is saddle if
 $\gamma\neq 2$ and repeller otherwise. It represents a solution completely dominated by the scalar field, and more specifically by its kinetic energy.

The second point, called $Y_+$, is located at $(x,y,w)=(0,1,0)$. Its eigenvalues are undefined because when the point is approached from different directions different results are obtained.  Numerical analysis suggests it is the late time attractor, it represents
a solution completely dominated by the  scalar field, and more specifically by its potential energy.

The third point, called $W_+$, is located at $(x,y,w)=(0,0,1)$. It is a saddle if $\gamma\ne0,2$, a repeller if $\gamma=2$ and it is undefined if $\gamma=0$. It represents a solution completely dominated by the fluid.

Finally the fourth point, called $\rm O$ is located at $(x,y,w)=(0,0,1)$. Its dynamical character is also undefined, but numerical analysis indicates it is a repeller. The choice of variables leading to the LELA system does not allow for an investigation of  this fixed point, but it must be noted that it comprises all the points in the HELA system except the 
 $\rm O$ point, so there is closes the circle, because the character of those points was investigated in the previous section.
 
 Note that the problems to define the dynamical character of some points in the HELA and the LELA systems should not come as a surprise, because this is a commonplace problem in
points lying outside the variety (points at which the equations or their derivatives blow up).

\vspace{0.1cm}
\begin{widetext}
\begin{center}
\begin{table}[h!]\caption{Eigenvalues and dynamical character  of the critical points of the
LELA dynamical system}
(assuming $n<1$, $0\le\gamma\le2$ and $\lambda^2\le1$)\vspace{0.3cm}\\
\begin{tabular}{ccc}
\hline \hline
Name &Eigenvalues&Dynamical Character\\
\hline \vspace*{5pt}
$X_{\pm}$ &\quad
$\left(-2(1-n),1,1-\displaystyle\frac{\gamma}{2}\right) $&\quad non-hyperbolic for $\gamma=2$, always repeller  (saddle for $0\le\gamma<2$)\quad
\\\\
$Y_{\pm}$ &
undefined&undefined
\\\\
$W_{\pm}$ &\quad $\left(-(1-n)\gamma,-1+\displaystyle\frac{\gamma}{2},\displaystyle
\frac{\gamma}{2}\right)$&\quad non-hyperbolic for $\gamma=0,2$, repeller for $\gamma>0$ (saddle for $0<\gamma<2$)
\\\\
$\rm O$ &$(0,0,0)$& undefined\\\vspace{-0.30cm}\\
\hline \hline
\end{tabular}
\vspace{20pt}\label{eigenB}
\end{table}
\end{center}
\end{widetext}

\vspace{2cm}
\section{Cosmological consequences}
The cosmological consequences of this analysis are simple but important.
As compared to the situation in standard cosmology, for the description of
Cardassian models we find that the  first complication stems from the necessity of  introducing an additional  variable, which we call $z$. \\

Our numerical analysis tell us that  the past attractors correspond to
$
\mathcal{S}(w,x,y)\equiv 1$, and because of the constraint the latter  
 enforces $z^{1-n}=0$ which  with in turn implies the recovery of the usual form of the Friedmann equation. In the case of  models expanding from an initial singularity will 
and for ($n<1$) we then conclude that the past attractors  corresponds more specifically to to $z=0$, and from the definition of $z$ we see that those are solutions with an initial singularity.  Summarizing, from the perspective of dynamical systems Cardassian models
with a fluid and a scalar field with an exponential potential will preferably have a big bang.

More specifically, the early-time attractor is a solution completely dominated by the kinetic energy of the scalar field and satisfying $ \rho\propto a^{-6}$, and its evolution is indistinguishable from that of perfect fluid models with a $p=(\gamma-1)\rho$ equation of state under the identification $n\equiv \gamma/2$, and the condition for inflation is simply $n<1/3$.

In contrast, the late-time  solution attractor is a solution completely dominated by the potential energy of the scalar field and satisfying $\rho\equiv\rm const.$, that is, the de Sitter solution, and since $H\equiv\rm const.$ too, inflation proceeds for any value of $n$.

Interestingly, there are no tracking late-time attractors, this is an important difference with respect to the behavior in standard (non-Cardassian) models.

\section{Conclusions}
Cardassian models have been proposed as yet one more possible explanation for late-time acceleration. The main interest of the proposal is it involves only matter and radiation and does not invoke either vacuum energy or a cosmological constant. The idea consists in introducing a modification to the Friedmann equation, so that the effects of the modification become important at low redshift.  

We have concentrated here on modifications of the form $3H^2-\rho\propto \rho^n$ with $n<1$,
and we have studied its asymptotic behavior assuming $\rho$ is made up of two contributions: the energy density of a perfect fluid with a $p=(\gamma-1)\rho$ equation of state and a self-interacting scalar field with an exponential potential.

Our analysis falls mainly on the analytical side, but we have also carried out some numerical investigations. We constructed two  dynamical system
arising from the evolution equations. The first system is best suited to telling us about
the asymptotic behavior in the high energy limit and we have called it our HELA system. The 
The second system, in contrast is best suited for a description of 
the asymptotic behavior in the low energy limit and we have called it our LELA system. 
We have found all the fixed points and we have characterized dynamically most of them by analytical means. However, the information regarding the dynamical character of some of the most important points seem to be only accessible by numerical methods. 

Our analysis allows us to say that the late-time attractor of the system is the de Sitter solution, and whatever the value of $n$, there will only be inflation. We have also found that the early-time attractor is a solution with a big-bang, and that there are not tracking late-time attractor. Basically, the behavior in the high energy regime is the same
as in standard cosmology but the modifications make the structure of late-time attractors rather different.

\section*{Acknowledgments}
Thanks to  J.M. Aguirregabiria and A. Gonzales for conversations. R.L. is supported by the Spanish Ministry of Science and Education through research grant FIS2004-01626, and G. Le\'on is supported by the MES of Cuba.

% ------------------------------------------------------------------------

\begin{thebibliography}{99}

\bibitem{supernovaefirst}
A.~G.~Riess {\it et al.}  [Supernova Search Team Collaboration],
% ``Observational Evidence from Supernovae for an Accelerating Universe and a
%Cosmological Constant,''
%
Astron.\ J.\  {\bf 116}, 1009 (1998);
%[arXiv:astro-ph/9805201].
%%CITATION = ASTRO-PH 9805201;%%
%\cite{Perlmutter:1998np}
%\bibitem{Perlmutter:1998np}
S.~Perlmutter {\it et al.}  [Supernova Cosmology Project Collaboration],
%``Measurements of Omega and Lambda from 42 High-Redshift Supernovae,''
Astrophys.\ J.\  {\bf 517}, 565 (1999);
%[arXiv:astro-ph/9812133].
%%CITATION = ASTRO-PH 9812133;%%
%\cite{Riess:2004nr}
%\bibitem{Riess:2004nr}

%\cite{Knop:2003iy}

R.~A.~Knop {\it et al.},
%``New Constraints on $\Omega_M$, $\Omega_\Lambda$, and w from an Independent
%Set of Eleven High-Redshift Supernovae Observed with HST,''
%
Astrophys.\ J.\  {\bf 598}, 102 (2003);
%[arXiv:astro-ph/0309368].
A.~G.~Riess {\it et al.}  [Supernova Search Team Collaboration],
%``Type Ia Supernova Discoveries at z>1 From the Hubble Space Telescope:
%Evidence for Past Deceleration and Constraints on Dark Energy Evolution,''
%
Astrophys.\ J.\  {\bf 607}, 665 (2004).
%[arXiv:astro-ph/0402512].
%%CITATION = ASTRO-PH 0402512;%%

\bibitem{wmap}
C.~L.~Bennett {\it et al.},
 %``First Year Wilkinson Microwave Anisotropy Probe (WMAP) Observations:
%Preliminary Maps and Basic Results,''
%
Astrophys.\ J.\ Suppl.\  {\bf 148}, 1 (2003);
%[arXiv:astro-ph/0302207].
%%CITATION = ASTRO-PH 0302207;%%
%\cite{Komatsu:2003fd}
%\bibitem{Komatsu:2003fd}
E.~Komatsu {\it et al.},
%``First Year Wilkinson Microwave Anisotropy Probe (WMAP) Observations: Tests
%of Gaussianity,''
%
Astrophys.\ J.\ Suppl.\  {\bf 148}, 119 (2003);
%[arXiv:astro-ph/0302223].
%%CITATION = ASTRO-PH 0302223;%%
%\cite{Hinshaw:2003ex}
%\bibitem{Hinshaw:2003ex}
G.~Hinshaw {\it et al.},
% ``First Year Wilkinson Microwave Anisotropy Probe (WMAP) Observations: Angular
%Power Spectrum,''
%
Astrophys.\ J.\ Suppl.\  {\bf 148}, 135 (2003);
%[arXiv:astro-ph/0302217].
%%CITATION = ASTRO-PH 0302217;%%
%\cite{Spergel:2003cb}
%\bibitem{Spergel:2003cb}
D.~N.~Spergel {\it et al.}  [WMAP Collaboration],
% ``First Year Wilkinson Microwave Anisotropy Probe (WMAP) Observations:
%Determination of Cosmological Parameters,''
%
Astrophys.\ J.\ Suppl.\  {\bf 148}, 175 (2003);
%[arXiv:astro-ph/0302209].
%%CITATION = ASTRO-PH 0302209;%%
%\cite{Peiris:2003ff}
%\bibitem{Peiris:2003ff}
H.~V.~Peiris {\it et al.},
% ``First year Wilkinson Microwave Anisotropy Probe (WMAP) observations:
%Implications for inflation,''
%
Astrophys.\ J.\ Suppl.\  {\bf 148}, 213 (2003).
%[arXiv:astro-ph/0302225].
%%CITATION = ASTRO-PH 0302225;%%
%\cite{Abazajian:2003jy}
\bibitem{sdss}
K.~Abazajian {\it et al.}  [SDSS Collaboration],
%``The First Data Release of the Sloan Digital Sky Survey,''
Astron.\ J.\  {\bf 126}, 2081 (2003),
%[arXiv:astro-ph/0305492].
%%CITATION = ASTRO-PH 0305492;%%
%\bibitem{Tegmark:2003ud}
M.~Tegmark {\it et al.}  [SDSS Collaboration],
%``Cosmological parameters from SDSS and WMAP,''
Phys.\ Rev.\ D {\bf 69}, 103501 (2004),
%[arXiv:astro-ph/0310723].
%%CITATION = ASTRO-PH 0310723;%%
%\cite{Tegmark:2003uf}
%\bibitem{Tegmark:2003uf}
M.~Tegmark {\it et al.}  [SDSS Collaboration],
%``The 3D power spectrum of galaxies from the SDSS,''
Astrophys.\ J.\  {\bf 606}, 702 (2004),
%[arXiv:astro-ph/0310725].
%%CITATION = ASTRO-PH 0310725;%%

%\cite{Abazajian:2004aj}
%\bibitem{Abazajian:2004aj}
K.~Abazajian {\it et al.}  [SDSS Collaboration],
%``The Second Data Release of the Sloan Digital Sky Survey,''
Astron.\ J.\  {\bf 128}, 502 (2004),
%[arXiv:astro-ph/0403325].
%%CITATION = ASTRO-PH 0403325;%%
%{Abazajian:2004it}
%\bibitem{Abazajian:2004it}
K.~Abazajian {\it et al.},
%``The Third Data Release of the Sloan Digital Sky Survey,''
arXiv:astro-ph/0410239.
%%CITATION = ASTRO-PH 0410239;%%

\bibitem{lewis} K. Freese and M. Lewis, Phys. Lett. B {\bf 540}, 1
(2002).

\bibitem{Netter} C.B Netterfield et al., Astrophys. J. {\bf 571},
604 (2002; R. Stompor et al., Astrophys. J. Lett. {\bf 561}, L7
(2001); N.W Halverson et al, Astrophys. J. {\bf 568}, 38 (2002);
C.Pryke et al., ibid. 568. 46 (2002)

\bibitem{chungfreese} D. J. Chung and K. Freese, Phys. Rev.  D {\bf
61}, 023511 (2000)

\bibitem{gondolofresse} P. Gondolo and K. Fresse,Phys. Rev. D{\bf
68}, 063509 (2003)

\bibitem{astro-ph/0201229} K. Freese and M. Lewis,
% "Cardassian Expanssion: a Model in which
%the Universe is Flat, Matter Dominated, and Accelerating."
Phys. Lett. B {\bf 540}, 1 (2002).

%\bibitem{64} C. Wetterich, Nucl. Phys. B {\bf 302}, 668 (1988)


%\bibitem{47} K. A. Olive, Phys. Rev. Rep. {\bf 190}, 307 (1990)

%\bibitem{48} A. Linde. Inflation and quantum cosmology, in 300
%Years of Gravitation, ed. S. W. Hawking and W. Israel, pp 604-630
%(Cambridge University Press, Cambridge,1987).

%\bibitem{50} A. H. Guth, Phys. Rev. D {\bf 23}, 347 (1981)

%\bibitem{49} A.P. Billyard, A.A. Coley and R.J. van den Hoogen, Phys.Rev. D 58, 123501 (1998) 


%\bibitem{65} D. Wands, E. J. Copeland, and A.R. Liddle, Ann. N. Y.
%Acad. Sci. {\bf 688}, 647 (1993)

%\bibitem{66} P. G. Ferreira and M. Joyce, Phys. Rev. D {\bf 58},
%023503 (1998).

\bibitem{wainlim} J. Wainwright and W. C. Lim
"Cosmological models from a dynamical systems perspective ", gr-qc/0409082.

\bibitem{copeland} E.~J.~Copeland, A.~R.~Liddle and D.~Wands,
%``Exponential potentials and cosmological scaling solutions,''
Phys.\ Rev.\ D {\bf 57}, 4686 (1998).



\bibitem{carrcoley} B. J. Carr and A. A. Coley, Class. Quantum
Grav. {\bf 16} R31-R71 (1999)


%\cite{Billyard:1998hv}
\bibitem{seminal}
A.~P.~Billyard, A.~A.~Coley and R.~J.~van den Hoogen,
%``The stability of cosmological scaling solutions,''
Phys.\ Rev.\ D {\bf 58}, 123501 (1998);
%%CITATION = GR-QC 9805085;%%
R.~J.~van den Hoogen, A.~A.~Coley and D.~Wands,
%``Scaling Solutions in Robertson-Walker Spacetimes,''
Class.\ Quant.\ Grav.\  {\bf 16}, 1843 (1999);
L.~P.~Chimento, A.~S.~Jakubi and D.~Pavon,
%``Enlarged Q-matter cosmology,''
Phys.\ Rev.\ D {\bf 62}, 063508 (2000);
A.~Nunes and J.~P.~Mimoso,
%``On the potentials yielding cosmological scaling solutions,''
Phys.\ Lett.\ B {\bf 488}, 423 (2000);
A.~Kehagias and G.~Kofinas,
%``Cosmology with exponential potentials,''
Class.\ Quant.\ Grav.\  {\bf 21}, 3871 (2004);
P.~J.~Steinhardt, L.~M.~Wang and I.~Zlatev,
%``Cosmological tracking solutions,''
Phys.\ Rev.\ D {\bf 59}, 123504 (1999);
L.~Amendola,
%``Scaling solutions in general non-minimal coupling theories,''
Phys.\ Rev.\ D {\bf 60}, 043501 (1999);
J.~P.~Uzan,
%``Cosmological scaling solutions of non-minimally coupled scalar fields,''
Phys.\ Rev.\ D {\bf 59}, 123510 (1999).
%%CITATION = GR-QC 9903004;%%


\bibitem{keilim} K.A. Olive, Phys. Rep. 190, 307 (1990);  Y. Kitada and M. Maeda, Class. Quan. Grav. {\bf 10}, 703 (1993).
\bibitem{wet} C. Wetterich, Nucl. Phys. B {\bf 302}, 668 (1988). 

\bibitem{exp} L.~P.~Chimento,
% ``General solution to two-scalar field cosmologies with exponential
%potentials,''
%
Class.\ Quant.\ Grav.\  {\bf 15}, 965 (1998); A.~R.~Liddle and R.~J.~Scherrer,
% ``A classification of scalar field potentials with cosmological scaling
%solutions,''
%
Phys.\ Rev.\ D {\bf 59}, 023509 (1999); A.~Nunes and J.~P.~Mimoso,
%``On the potentials yielding cosmological scaling solutions,''
Phys.\ Lett.\ B {\bf 488}, 423 (2000);A.~Kehagias and G.~Kofinas,
%``Cosmology with exponential potentials,''
Class.\ Quant.\ Grav.\  {\bf 21}, 3871 (2004).


\bibitem{coptrack} E.J. Copeland, M.R. Garousi, M. Sami, and S. Tsujikawa, Phys.\ Rev.\ D {\bf 71}, 043003 (2005).
\bibitem{hirsch} M. Hirsch and S. Smale, {\it Differential Equations, Dynamical Systems, and Linear Algebra}, 2nd ed. (Academic Press, New York, 1974).
\bibitem{supernovae}
Y.~Wang, K.~Freese, P.~Gondolo and M.~Lewis,
  %``Future type IA supernova data as tests of dark energy from modified
  %Friedmann equations,''
  Astrophys.\ J.\  {\bf 594}, 25 (2003);
   Z.~H.~Zhu, M.~K.~Fujimoto and X.~T.~He,
  %``Observational constraints on cosmology from modified Friedmann equation,''
  Astrophys.\ J.\  {\bf 603}, 365 (2004);
M.~Szydlowski and W.~Czaja,
  %``Toward reconstruction of the dynamics of the Universe from distant type Ia
  %supernovae,''
  Phys.\ Rev.\ D {\bf 69} (2004) 083507.
\bibitem{phantom}
R.~R.~Caldwell,
%``A Phantom Menace?,''
Phys.\ Lett.\ B {\bf 545}(2002) 23 ;
%\cite{Gonzalez-Diaz:2003rf}
P.~F.~Gonzalez-Diaz,
%``K-essential phantom energy: Doomsday around the corner?,''
Phys.\ Lett.\ B {\bf 586}(2004) 1;
P.~Singh, M.~Sami and N.~Dadhich,
%``Cosmological dynamics of phantom field,''
Phys.\ Rev.\ D {\bf 68} (2003) 023522;
%%CITATION = HEP-TH 0305110;
G.~Calcagni,
%``Regularized dualities in patch cosmology,''
arXiv:gr-qc/0410111; V.~B.~Johri,
%``Phantom cosmologies,''
Phys.\ Rev.\ D {\bf 70} (2004) 041303;
V.K.Onemli and R.P.Woodard, Class. Quant. Grav.{\bf19} (2000) 4607
; S. Hannestad and E. Mortsell, Phys. Rev.
D{\bf66},063508(2002); S.~M.~Carroll, M.~Hoffman and M.~Trodden,
%``Can the dark energy equation-of-state parameter w be less than -1?,''
Phys.\ Rev.\ D {\bf 68} (2003) 023509;
P.H.Frampton, hep-th/0302007; P.~F.~Gonzalez-Diaz and
C.~L.~Siguenza, %``Phantom thermodynamics,''
Nucl.\ Phys.\ B {\bf 697} (2004) 363;
G.W.Gibbons, hep-th/0302199; B.Mcinnes, astro-ph/0210321;
L.~P.~Chimento and R.~Lazkoz, Phys.\ Rev.\ Lett.\  {\bf 91}
 (2003) 211301; H.~Stefancic,Phys.\ Lett.\ B
{\bf 586} (2004) 5; H.~Stefancic, Eur.\
Phys.\ J.\ C {\bf 36} (2004) 523;
P.~F.~Gonzalez-Diaz, Phys.\ Rev.\ D {\bf 69}, 063522
(2004); V.~K.~Onemli and R.~P.~Woodard,
%``Quantum effects can render w < -1 on cosmological scales,''
Phys.\ Rev.\ D {\bf 70} (2004)  107301;
T.~Brunier, V.~K.~Onemli and R.~P.~Woodard,
%``Two loop scalar self-mass during inflation,''
Class.\ Quant.\ Grav.\  {\bf 22} (2005) 59;
E.~Elizalde, S.~Nojiri and S.~D.~Odintsov,
%``Late-time cosmology in (phantom) scalar-tensor theory: Dark energy and the
%cosmic speed-up,''
Phys.\ Rev.\ D {\bf 70} (2004) 043539  [arXiv:hep-th/0405034];
S.~Nojiri and S.~D.~Odintsov,
%``Effective equation of state and energy conditions in phantom / tachyon
%inflationary cosmology perturbed by quantum effects,''
Phys.\ Lett.\ B {\bf 571} (2003) 1;  %[arXiv:hep-th/0306212];
S.~Nojiri and S.~D.~Odintsov,
%``Quantum deSitter cosmology and phantom matter,''
Phys.\ Lett.\ B {\bf 562} (2003) 147.  %[arXiv:hep-th/0303117].
\bibitem{olkit}K.A. Olive, Phys. Rep. 190 (1990) 307, Y. Kitada and M. Maeda, Class. Quantum Grav. 10 (1993) 703.
\bibitem{favor}Y.~Wang, K.~Freese, P.~Gondolo and M.~Lewis,
  %``Future type IA supernova data as tests of dark energy from modified
  %Friedmann equations,''
  Astrophys.\ J.\  {\bf 594} (2003) 25; Z.~H.~Zhu, M.~K.~Fujimoto and X.~T.~He,
  %``Observational constraints on cosmology from modified Friedmann equation,''
  Astrophys.\ J.\  {\bf 603}, 365 (2004); S.~Nesseris and L.~Perivolaropoulos,
  %``A comparison of cosmological models using recent supernova data,''
  Phys.\ Rev.\ D {\bf 70} (2004) 043531;
R.~Lazkoz, S.~Nesseris and L.~Perivolaropoulos,
  %``Evidence for Cosmological Oscillations in the Gold SnIa Dataset,''
  astro-ph/0503230.


\bibitem{cardin}
 M.~Szydlowski and W.~Czaja,
  %``Modified Friedmann Cosmologies - theory \& observations,''
  astro-ph/0402510.
\end{thebibliography}
\end{document}